\documentclass[10pt,superscriptaddress]{revtex4}
\pdfoutput=1
\usepackage{amsmath,amssymb,amsfonts,epsfig,graphicx,euscript}%
\usepackage[pdftex,bookmarks,bookmarksnumbered,linktocpage,pdfstartview=FitH]{hyperref}
\hypersetup{colorlinks,%
citecolor=red,%
filecolor=blue,%
linkcolor=blue,%
urlcolor=blue,%
pdftex}
\usepackage[all]{hypcap}
\numberwithin{equation}{section}
\newcommand{\bse}{\begin{subequations}}
\newcommand{\ese}{\end{subequations}}
\newcommand{\be}{\begin{equation}}
\newcommand{\ee}{\end{equation}}
\newcommand{\bea}{\begin{eqnarray}}
\newcommand{\eea}{\end{eqnarray}}
\newcommand{\ba}{\begin{array}}
\newcommand{\ea}{\end{array}}

\begin{document}
\hfill%
\vbox{
    \halign{#\hfil        \cr
           IPM/P-2014/057\cr
                     }
      }
\vspace{0.5cm}
\begin{center}
{ \Large{\textbf{Time-Dependent Meson Melting in External Magnetic Field}}} %\\
\vspace*{1.2cm}
\begin{center}
{\bf M. Ali-Akbari$^{a,b,1}$, F. Charmchi$^{b,2}$,
A. Davody$^{b,3}$, H. Ebrahim$^{c,d,4}$, L. Shahkarami$^{e,5}$}\\%
%{\bf M. Ali-Akbari$^{a,b,}$\footnote{aliakbari@theory.ipm.ac.ir}, F. Charmchi$^{b}$\footnote{aaa},
%A. Davody$^{b}$\footnote{davody@ipm.ir}, H. Ebrahim$^{c,d}$\footnote{hebrahim@ipm.ir}, L. Shahkarami$^{e}$\footnote{aaa}}\\%
\vspace*{0.3cm}
{\it {${}^a$Department of Physics, Shahid Beheshti University G.C., Evin, Tehran 19839, Iran}}\\
{\it {${}^b$School of Particles and Accelerators, Institute for Research in Fundamental Sciences (IPM),
P.O.Box 19395-5531, Tehran, Iran}}  \\
{\it {${}^c$Department of Physics, University of Tehran, North Karegar Ave., Tehran 14395-547, Iran}}\\
{\it {${}^d$School of Physics, Institute for Research in Fundamental Sciences (IPM),
P.O.Box 19395-5531, Tehran, Iran}}  \\
{\it {${}^e$School of Physics, Damghan University, Damghan, 41167-36716, Iran}}\\
{\it {${}^1$ $m_{-}$aliakbari@sbu.ac.ir}, {${}^2$charmchi@ipm.ir},
 {${}^3$davody@ipm.ir}, {${}^4$hebrahim@ut.ac.ir}, {${}^5$l.shahkarami@du.ac.ir}   }

\vspace*{0.5cm}
\end{center}
\end{center}

%\vspace{.2cm}
\bigskip
\begin{center}
\textbf{Abstract}
\end{center}
The dynamics of a probe D7-brane in an asymptotically AdS-Vaidya background has been investigated in the presence of an external magnetic field. Holographically, this is dual to the dynamical meson melting in the ${\cal{N}}=2$ super Yang-Milles theory. If the final temperature of the system is large enough, the probe D7-brane will dynamically cross the horizon (black hole embedding). By turning on the external magnetic field and raising it sufficiently, the final embedding of the corresponding D7-brane changes to Minkowski embedding. In the field theory side, this means that the mesons which melt due to the raise in the temperature, will form bound states again by applying an external magnetic field. We will also show that the evolution of the system to its final equilibrium state is postponed due to the presence of the magnetic field.
%\newpage
\vspace*{0.7cm}

\tableofcontents

\section{Introduction}
A new phase of QCD, called quark-gluon plasma (QGP), is produced at RHIC and LHC by colliding two heavy nuclei such as gold or lead, respectively, at relativistic speeds \cite{Shuryak:2003xe}. Studying the properties of the QGP has attracted a lot of attention. For instance, the process of plasma formation and equilibration has been considerably investigated over the last couple of years. In fact it is generally agreed that the plasma thermalizes very rapidly. By thermalization one usually means that the plasma reaches the stage after which the hydrodynamic equations describe the evolution of the system, . Experimental observations indicate that such a plasma is strongly coupled and numerical simulations suggest that at the early stages of the plasma formation a strong magnetic field is produced \cite{Kharzeev:2007jp}. The effect of the presence of a strong magnetic field on the properties of the plasma has been studied extensively in the literature \cite{Erdmenger, AliAkbari:2012vt}.

Strongly coupled nature of the QGP makes perturbative methods in studying its properties not applicable. A reasonable candidate to overcome this problem is AdS/CFT correspondence or more generally gauge/gravity duality. This duality states that a $d$-dimensional strongly coupled Yang-Mills theory is dual to a $(d+1)$-dimensional classical gravity theory on asymptotically AdS space-time \cite{Maldacena}. The corresponding field theory lives on the boundary of the dual theory (bulk gravity).

In the AdS/CFT dictionary the vacuum and an arbitrary thermal state of the field theory correspond to pure AdS and AdS black hole, respectively. Thus the thermalization process in the field theory can be equivalent to the black hole formation in the bulk as has been shown in \cite{Chesler:2008hg}. For instance the Vaidya metric which is the result of the collapse of the matter in the bulk can describe the dynamics of the plasma from zero to a finite temperature.

The black hole formation in the bulk resembles thermalization in the gluon sector of the field theory. One can also study such a process in the meson sector, called {\it{meson melting}}. Depending on the temperature of the plasma, some of the quarkonium mesons (heavy mesons) stay stable and live in the plasma.  The injection of enough energy can cause these mesons to  melt. Holographically, mesons are described by the fluctuations of the shape of the probe D-branes in the asymptotically AdS backgrounds \cite{Karch:2002sh}. The mesons are stable if the probe brane does not touch the bulk horizon and they melt when it crosses the horizon \cite{Mateos}. The dynamical meson melting by which we mean the process of the meson melting in a time-dependent background (Vaidya metric) has been investigated in \cite{Ishii:2014paa}. In this paper we are interested in studying the effect of a non-zero external magnetic field on the dynamics of the meson melting. We will also discuss how the magnetic field affects the equilibration time when the system gets to its final equilibrium situation.

\section{Probe D7-brane in the Vaidya Background}
In this section we will study the set-up of the dynamical meson melting in the presence of a constant external magnetic field. According to the AdS/CFT dictionary the mesons in the gauge theory correspond to the fluctuations of the probe brane in the asymptotically AdS background. In fact, in the AdS-black hole background, the shape of the brane can be categorized into Minkowski Embedding (ME)  and Black Hole Embedding (BE) which correspond to the mesonic (stable mesons) and melted (unstable mesons) phases, respectively \cite{Mateos}. The process of dynamical meson melting is described by the change in the shape of the probe brane dynamically. In order to achieve this goal we study the shape of the probe brane in the AdS-Vaidya background. It describes the black hole formation in the bulk which corresponds to thermalization in the gluon sector of the dual gauge theory. The background is time-dependent and therefore the shape of the brane in such a background can deform dynamically.

Let us start with the AdS-Vaidya metric that is
\be%
ds^2 =G_{MN} dx^M dx^N= \frac{1}{z^2} \left[-F(V,z) dV^2 - 2 dV dz + d{\vec{x}}_3^2 \right] + d\phi^2+\cos^2\phi ~d\Omega_3^2+\sin^2\phi ~d\psi^2,
\ee%
where
\be%
F(V,z) = 1-M(V) z^4,
\ee%
and we have set the radius of AdS space-time to be one, $R=1$. The above metric is written in Eddington-Finkelstein coordinate where the radial direction is represented by $z$ and $V$ shows the null direction. The boundary, where the gauge theory lives, is at $z=0$ and $V$ is the time coordinate on the boundary. $M(V)$, which is an arbitrary function, represents the mass of the black hole which changes as time passes by until it reaches a constant value. The function for $M(V)$ that we will work with in this paper, is
\bea %
\label{M}
 M(V)= M_f \left\{%
\begin{array}{ll}
    0 & V<0, \\
    \frac{1}{2}\left[1-\cos(\frac{\pi V}{\Delta V})\right] & 0 \leqslant V \leqslant \Delta V, \\
   1 & V>\Delta V ,\\
\end{array}%
\right. \eea %
where $\Delta V$ is the time interval in which the mass of the black hole increases from zero to $M_f$ which is constant. Note that the radius of the event horizon is $r_h=M_f^{\frac{1}{4}}$.

In order to add the fundamental matter in the gauge theory side we have to add the probe branes to the bulk. The dynamics of the degrees of freedom living on the brane is explained by the DBI action as
\be %
 S=-\tau_7\int d^8 \xi\sqrt{-g_{ab}+(2\pi\alpha')F_{ab}}~,
\ee %
where $g_{ab}$ is the induced metric and $F_{ab}$ is the gauge field strength on the brane. $a,\, b$ are the brane coordinates and $g_{ab}$ is defined as
\be%
g_{ab} = \partial_a x^M  \partial_b x^N G_{MN}.
\ee%
We are following the same notations as ones introduced in \cite{Ishii:2014paa} and will not repeat the details. The D7-brane is embedded along the six directions of the bulk metric, $\vec{x}$ and $\Omega_3$. We choose the other two coordinates on the brane to be null coordinates $u$ and $v$. For the rest of the bulk coordinates we choose the ansatz
\be %
 V=V(u,v),\ \ z=Z(u,v),\ \ \phi=\Phi(u,v),\ \ \psi=0.
\ee %
Since we are interested in studying the effect of the external magnetic field on the dynamical meson melting, we choose the magnetic field to be
\be%
F_{x_1 x_2} = B,
\ee%
that is constant.

Since we would like the $u$ and $v$ coordinates to stay null, we have to impose the following constraint equations on the equations of motion for the fields on the brane, obtained from the DBI action
\bse\label{cons}\begin{align}
 FV_{,v}^2+2Z_{,v}V_{,v}-Z^2(Z\Psi)^2_{,v}&=0, \\
 FV_{,u}^2+2Z_{,u}V_{,u}-Z^2(Z\Psi)^2_{,u}&=0,
\end{align}\ese
where $\Psi(u,v) \equiv \frac{\Phi(u,v)}{Z(u,v)}$ which, in fact, gives the shape of the brane. With the above assumptions the DBI action becomes
\be%
S=-\tau_7 V_{\Omega_3} V_{\vec{x}} \int du dv \frac{\cos^3 \Phi}{Z^3} \sqrt{\frac{1}{Z^4} + (2 \pi \alpha')^2 B^2} \left(F V_{,u} V_{,v} + V_{,u} Z_{,v} +Z_{,u} V_{,v} - Z^2 \Phi_{,u} \Phi_{,v}\right).
\ee%
The difference between the above action and the one in \cite{Ishii:2014paa} is that we have considered non-zero magnetic field which appears in the square root. Therefore, the equations of motion read
\bse\begin{align}
V_{,uv}&=
 \frac{Z{\cal{B}}_1^+}{2}  (Z\Psi)_{,u}(Z\Psi)_{,v}
+\frac32 \tan(Z\Psi)[(Z\Psi)_{,u}V_{,v}+(Z\Psi)_{,v}V_{,u}]
+\frac{1}{2}\left(F_{,Z}-\frac{F{\cal{B}}_2}{Z} \right)V_{,u}V_{,v}
\, ,
\\
Z_{,uv}&=
-\frac{ F Z {\cal{B}}_1^+}{2} (Z\Psi)_{,u}(Z\Psi)_{,v}
+ \frac32 \tan(Z\Psi)[(Z\Psi)_{,u}Z_{,v}+(Z\Psi)_{,v}Z_{,u}]+\frac{{\cal{B}}_2}{Z} Z_{,u}Z_{,v}- \frac{F_{,V}}{2}V_{,u}V_{,v}\nonumber
\\
&- \frac{1}{2}\left(
 F_{,Z}-\frac{F {\cal{B}}_2}{Z}
\right)(FV_{,u}V_{,v}+V_{,u}Z_{,v}+V_{,v}Z_{,u})
\, ,\\
\Psi_{,uv}&=
\left(\frac{3\tan(Z\Psi)}{2Z}+ \frac{\Psi F {\cal{B}}_1^+}{2} \right)(Z\Psi)_{,u}(Z\Psi)_{,v}
- \frac{{\cal{B}}_1^- + 3 Z \Psi \tan(Z\Psi) }{2Z^2}
[(Z\Psi)_{,u}Z_{,v}+(Z\Psi)_{,v}Z_{,u}]\nonumber\\
&+ \frac{\Psi}{2Z}\left(
 F_{,Z}-\frac{F{\cal{B}}_2}{Z} + \frac{3\tan(Z\Psi)}{Z^2\Psi}
\right)
(FV_{,u}V_{,v}+V_{,u}Z_{,v}+V_{,v}Z_{,u})
- \frac{\Psi {\cal{B}}_1^+}{Z^2}Z_{,u}Z_{,v}
+ \frac{\Psi F_{,V}}{2Z}V_{,u}V_{,v}\, ,
\
\end{align}\ese
where
\bse\begin{align} %
 {\cal{B}}_1^\pm&=1 \pm \frac{2}{1+(2\pi\alpha')^2 Z^4 B^2},\\
 {\cal{B}}_2&=3+\frac{2}{1+(2\pi\alpha')^2 Z^4 B^2}.
\end{align}\ese %
By setting $B=0$ we can recover the results of \cite{Ishii:2014paa}.

The above equations of motion are solvable, in spite of being non-linear and time-dependent, if one uses numerical methods. Here we apply finite difference method as has been used in \cite{Ishii:2014paa}. The solution to these equations will tell us how the shape of the brane changes due to the dynamical background and the presence of the non-zero magnetic field. According to the AdS/CFT dictionary the near boundary expansion of the shape of the brane gives the mass of quark $(m)$ and the quark condensate $(c)$. More specifically
\be%
\Psi(V,Z)|_{Z\rightarrow 0} = m(V) + \left( c(V) + \frac{m(V)^3}{6} \right) Z^2 + ....
\ee%
Note that in the presence of the magnetic field we rescale all the parameters with $m$, for instance $B$ is rescaled as $\frac{B}{m^2}$. This means that $m=1$ in all our results. By solving the equations of motion we will be able to see how the quark condensate behaves in time. The boundary and initial conditions are summarized in the following subsection.

\subsection{Boundary conditions and initial data}
In order to solve the equations of motion we need to define the appropriate boundary and initial conditions. To begin we have to distinguish between two configurations mentioned before; Minkowski embedding and Black Hole embedding. Since the ME does not cross the horizon, we have to impose two boundary conditions; one at the AdS boundary $Z|_{u=v}=0$, where we have used the same definition as \cite{Ishii:2014paa}, and the other at the pole $\Phi|_{u=v+\frac{\pi}{2}}=\frac{\pi}{2}$, where the 3-sphere on which the D7-brane wraps, shrinks to zero. For the BE, the 3-sphere never shrinks to zero as the D7-brane crosses the horizon and therefore we do not need to impose any boundary condition at the pole. In order to derive the appropriate boundary and initial conditions in the presence of the magnetic field, we follow the same path as \cite{Ishii:2014paa} where we refer the reader to it for more details.

\begin{itemize}
\item {\bf{Boundary conditions at the AdS boundary}}

The physical observables are obtained on the AdS boundary, where $u=v$ and therefore we have to include the appropriate boundary conditions there. The boundary condition on $Z$ and $\Phi$ are simply $Z|_{u=v}=0$ and $\Psi|_{u=v}=m$. In order to obtain the rest of the boundary conditions we expand the fields near the boundary and by imposing the regularity condition on the equations of motion and consistency with the constraint equations \eqref{cons} we get
\be%
\frac{d}{dv} V_0(v) = 2 Z_{,u}|_{u=v}~,~~~Z_{,uv}|_{u=v}=0~,
\ee%
where $V_0(v)=V|_{u=v}$. Note that the boundary conditions are not affected by the presence of the external magnetic field.

\item  {\bf{Boundary conditions at the Pole}}

Similar to the previous set of boundary conditions the presence of the magnetic field does not alter the boundary conditions at the pole, $\Phi|_{u=v+\frac{\pi}{2}}=\frac{\pi}{2}$ for vanishing magnetic field. Therefore we have
\be%
Z_{,u}=Z_{,v}~,~~~~V_{,u}=V_{,v}~,
\ee%
at $u=v+\frac{\pi}{2}$.

\item {\bf{Initial data}}

The background at which the probe brane is embedded is time-dependent where before the energy injection, $V<0$, the space-time is pure AdS. Therefore, our initial data includes the static solution to the equations of motion for pure AdS with non-zero magnetic field \cite{Erdmenger, AliAkbari:2012vt}.
\end{itemize}

\section{Numerical Results}
In this section we study the response of the system to the time-dependent change in the temperature in the presence of an external magnetic field. The response of the system is described by the behaviour of quark condensate, $c$, in terms of the boundary time. When the magnetic field is zero, this behaviour can be classified into three categories: Minkowski embedding, Black Hole embedding and Overeager case.
\begin{figure}
\begin{center}
  \includegraphics[width=150mm]{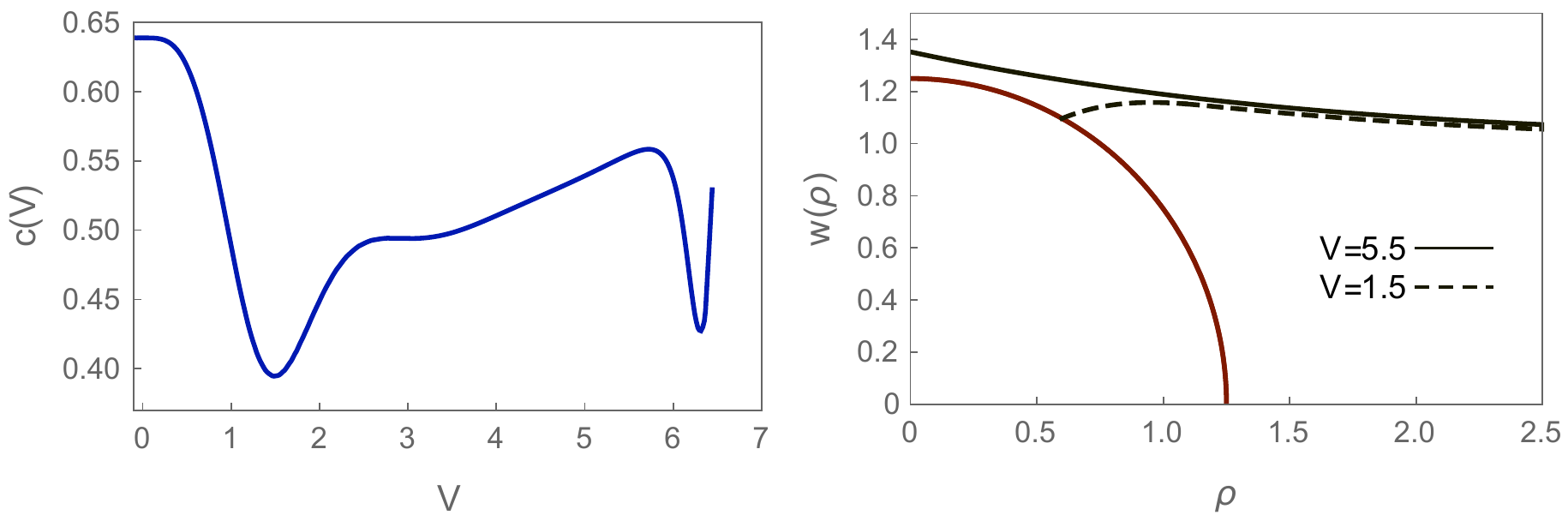}
\caption{This figure (right) shows that how an overeager configuration crosses the horizon and jumps out of it and becomes ME. $W(\rho)=z^{-1}\sin\phi$ reperesents the shape of the brane \cite{Ishii:2014paa}. Accordingly, the graph on the left shows that $c(V)$ starts oscillating at the final stages, as it is expected from ME. This figure is plotted for $B=1.87140$, $\Delta V=0.5$ and $r_h=1.25$.}\label{overeager2}
\end{center}
\end{figure}

\begin{itemize}
\item {\bf{Minkowski Embedding}}

For this case one starts from an initial ME at zero temperature and by increasing the temperature in the bulk the final configuration is still ME. Since the initial ME state represents the mass of the quark in the dual field theory, if the ratio of $m$ to the final temperature of the system is much larger than one, we can be sure that the embedding will always stay Minkowski. Therefore by raising the temperature, the phase transition does not happen and the mesons survive in the system. In this case the quark condensate shows an oscillatory behaviour in time about its equilibrium value.

\item {\bf{Black Hole Embedding}}

If $\frac{m}{T}$ becomes smaller than one, even though the initial state is ME, the D7-brane crosses the horizon as the temperature is increased and thus the final configuration becomes BE. In this situation the quark condensate falls from the initial zero value to its final equilibrium amount.

\item {\bf{Overeager Case}}

Provided that the time-scale of the change in the temperature is small, if the corresponding mass of the initial ME configuration is larger but close to the final temperature, the overeager case happens. This means that the D7-brane crosses the horizon and returns back again to another ME. In this case $c(V)$ decreases from zero to a minimum value and then starts increasing to its final equilibrium amount.
\end{itemize}
\begin{figure}
\begin{center}
  \includegraphics[width=70mm]{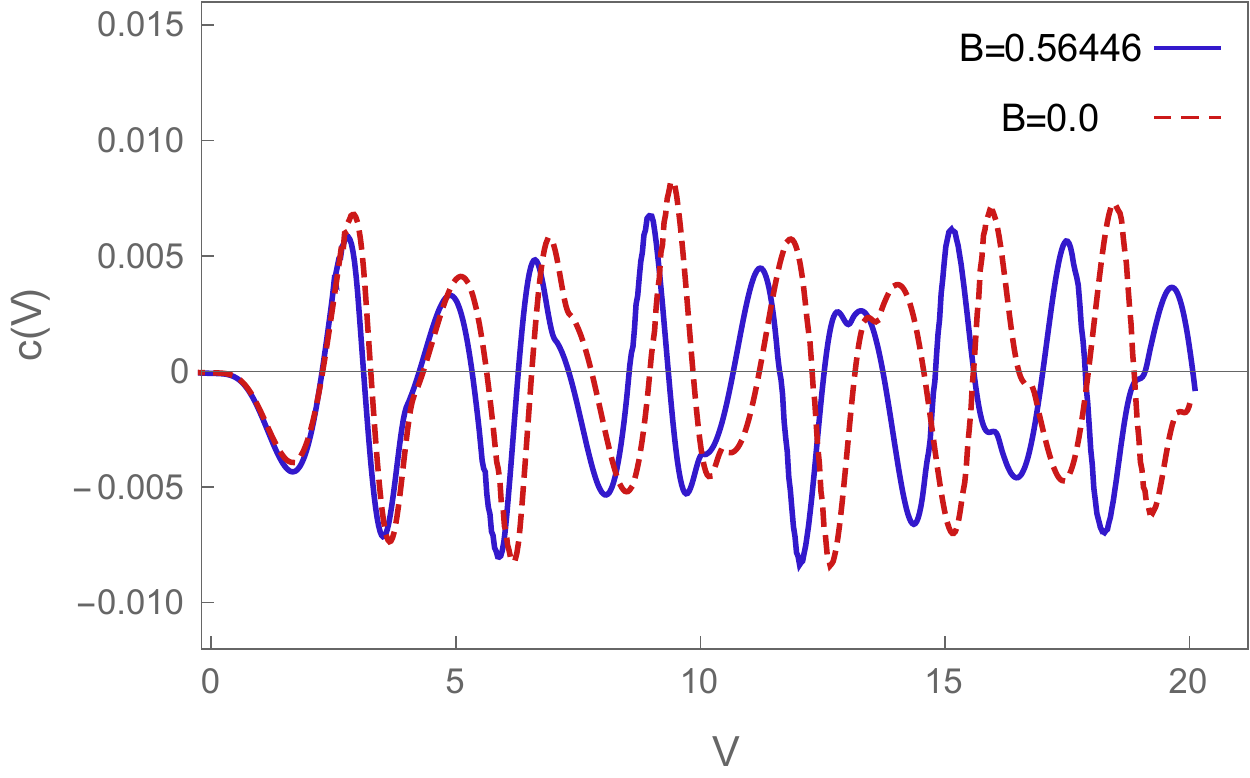}
  \hspace{5mm}
  \includegraphics[width=70mm]{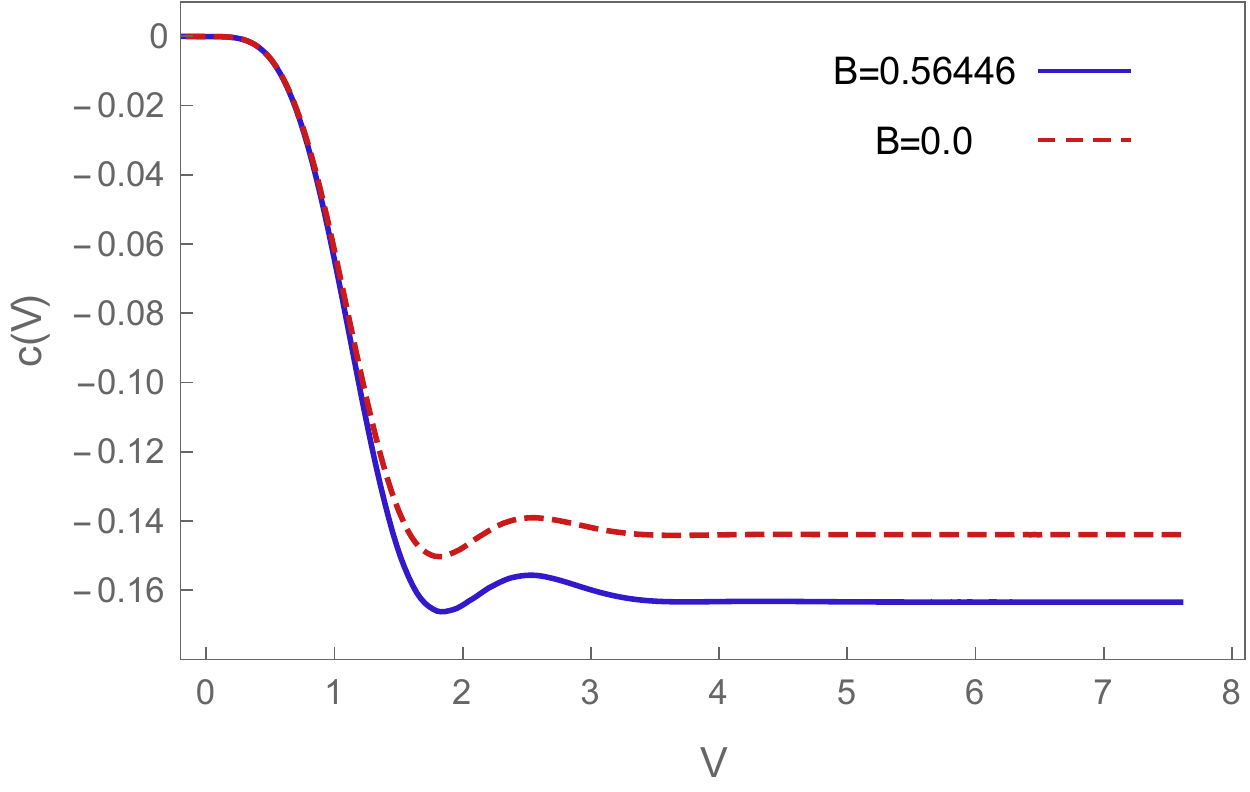}
\caption{$c(V)$ for ME (left) and BE (right). In both figures blue solid curves show $c(V)-c_0$, with B=0.56446, in which $c_0=0.0747132$. $c_0$ is the initial configuration quark condensate, before the black hole formation in the bulk. The red dashed curves show $c(V)$ in the absence of magnetic field. In all cases $r_h=1.25$ and $\Delta V = 1.0$.  }\label{C(V)ME}
\end{center}
\end{figure}
Before reporting our results for the non-zero magnetic field we would like to mention an interesting observation in the overeager case. As can be seen in the figure \ref{overeager2}, $c(V)$ starts oscillating at the final stages of its time-dependent evolution. This oscillation is around a final equilibrium value which in fact is the $c$ for a ME in the presence of the corresponding black hole in static case. This also confirms that the overeager solution ends up in ME. Looking at the plot of the shape of the brane, figure \ref{overeager2} (right), we can see that at the initial times after the energy injection the shape of the brane is described by the BE while at sufficiently later times it becomes ME.

We have studied the effect of the presence of a non-zero external magnetic field on the above embeddings as presented in figure \ref{C(V)ME}. Figure \ref{C(V)ME} left(right) compares the behaviour of the ME(BE) between the two cases. Note that the blue curves for non-zero $B$ are shifted in order to have the starting point of $c(V)$ plots at $c=0$. As observed in the figure, the equilibrium value of $c$ decreases for $B\ne 0$ and the oscillations happen with a smaller phase velocity.

\begin{figure}
\begin{center}
 \includegraphics[width=80mm]{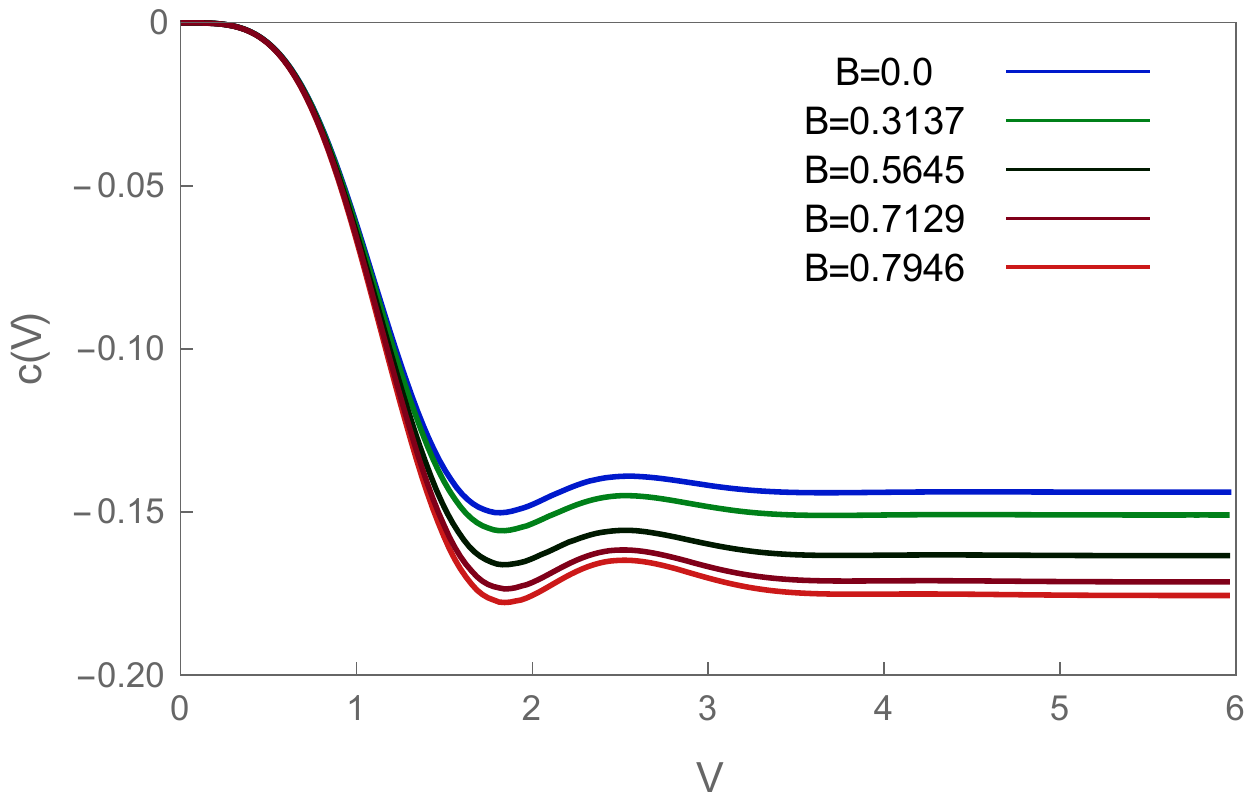}
 \hspace{.5cm}
  \includegraphics[width=80mm]{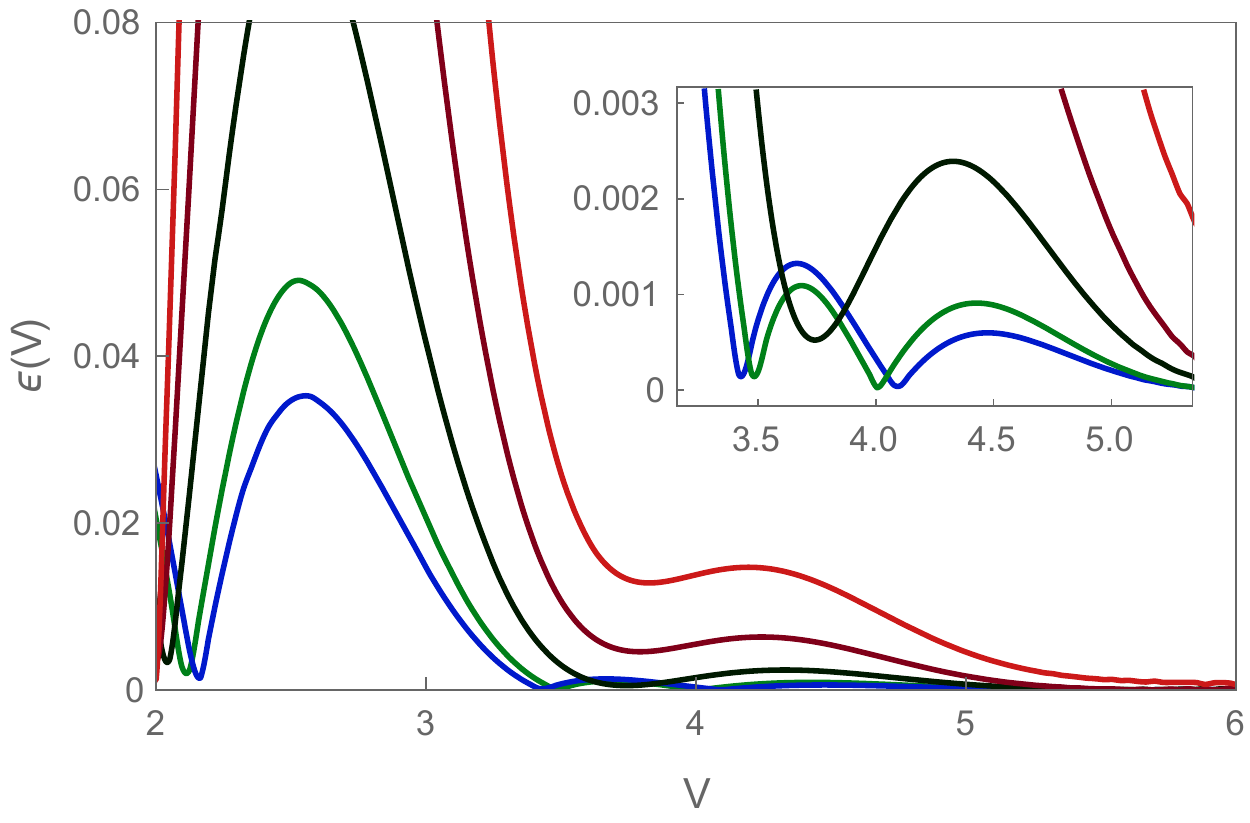}
\caption{The plot on the left shows $c(V)$ for BEs for different values of $B$. The graph on the right shows $\epsilon(V)$ with respect to the boundary time, obtained from \eqref{equ}, for the corresponding plots in the left figure. The equilibration times are represented in table \ref{tableteq}. We have set $r_h = 1.25$ and $\Delta V = 1.0$.}\label{thtime}
\end{center}
\end{figure}

\begin{table}[ht]
\label{table}
\caption{Bound-States or Stable Mesons obtained from the ME Power Spectrum}
\vspace{.3cm}
\centering
\begin{tabular}{c c c }
\hline\hline
Stable Modes ~~   &   ~~ $B=0$ ~~& ~~ $B=0.56446$   \\[0.5ex]
\hline
$\omega_1$ & 2.82052 &  3.04731\\
$\omega_2$ & 4.83518 &  5.18042\\
$\omega_3$ & 6.84984 &  7.21553 \\[1ex]

\hline
\end{tabular}
\end{table}
We have computed the power spectrum for the ME oscillations in both cases and present the results in table \ref{table}. The power spectrum is obtained from the discrete Fourier transform of $c(V)$. In fact the peaks of the power spectrum give us the mass spectrum of the stable mesons. Interestingly we observe that the meson masses increases by raising the magnetic field as seen in table \ref{table}. Since for the binding energy of the stable mesons we have $E_B\sim \sqrt{\lambda} M_{mes}$ \cite{CasalderreySolana:2011us}, where $\lambda$ is the t'Hooft coupling constant, the binding energy increases in the presence of the magnetic field. We should emphasize that the results obtained here come form solving the non-linear equations of motion rather than considering the linear perturbation around the equilibrium shape of the brane. Figure \ref{C(V)ME} (right) confirms the fact that $c(V)$ decreases by applying an external magnetic field. We will elaborate more on this in figure \ref{thtime}.

\begin{figure}
\begin{center}
  \includegraphics[width=130mm]{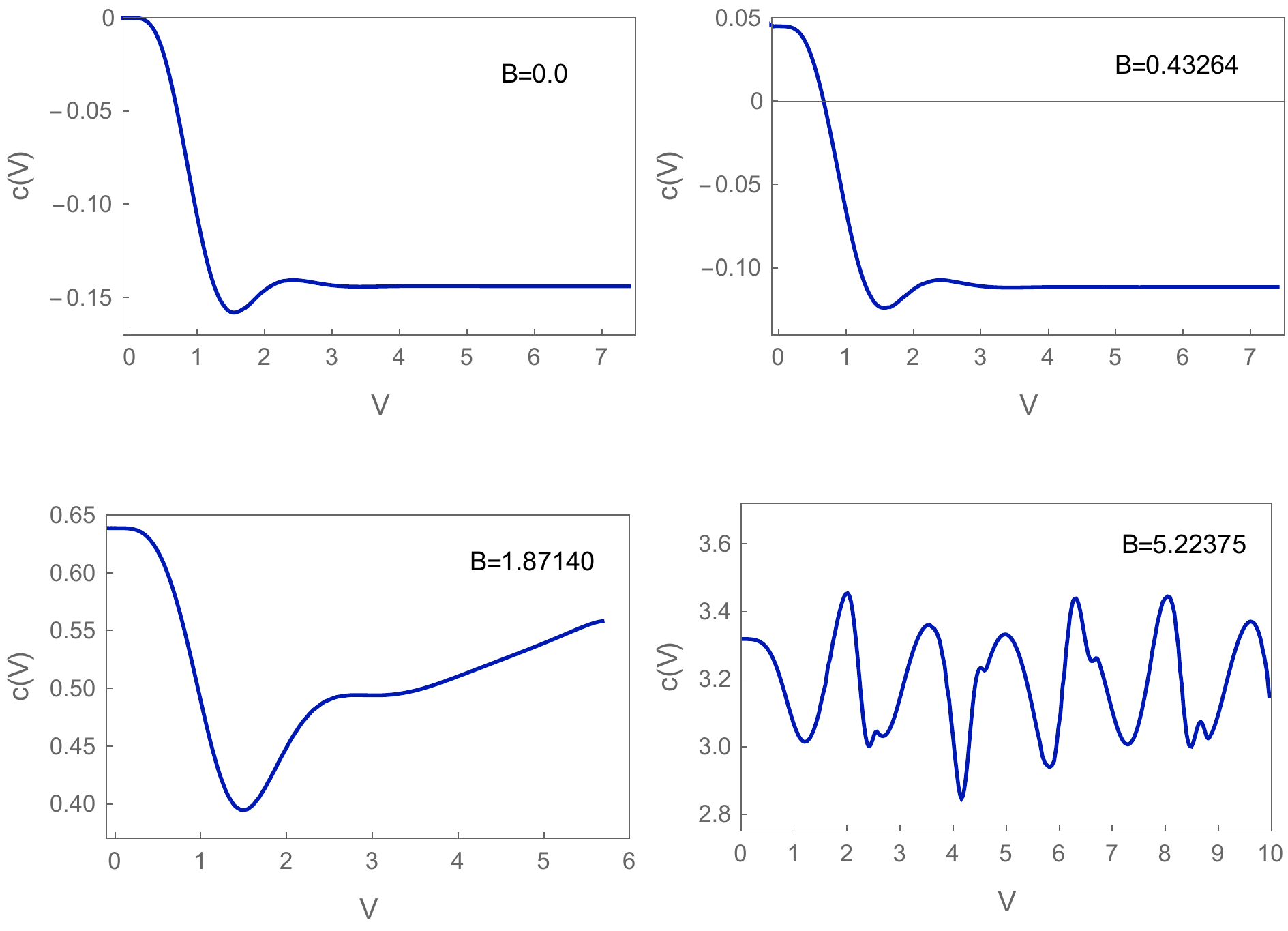}
\caption{This figure shows how the BE changes to overeager and consequently to ME after one raises the magnetic field. We have set $r_h=1.25$ and $\Delta V=0.5$.}\label{overeager}
\end{center}
\end{figure}

A remarkable phenomenon has been observed in the presence of the external magnetic field. Let us assume we start from a ME which results in BE after the injection of energy when the magnetic field is zero, figure \ref{overeager} up-left. If we raise the magnetic field, it can still give BE as can be seen in figure \ref{overeager} up-right. Interestingly, by further increasing the value of the magnetic field, with the same initial configuration and time-scale of the energy injection, the final configuration becomes overeager, figure \ref{overeager} down-left. If the magnetic field becomes large enough, the configuration stays ME and never touches the horizon, as shown in figure \ref{overeager} down-right. It is instructive to see what this means in the field theory side. At zero magnetic field, due to the energy injection the mesons are melted; indicating that quarks and anti-quarks live freely in the plasma. Raising the external magnetic field causes them to form bound-states again and the mesons become stable. This is consistent with our physical intuition from the non-dynamical picture. As has been discussed in \cite{AliAkbari:2012vt}, this observation can be explained by the fact that due to the non-zero magnetic field the probe brane becomes resistant to change in its shape. In the field theory side this means that the binding energy of the mesons become larger and therefore, more energy is needed to melt the mesons.

When we start from a ME which, due to the energy injection, ends up in a BE, there exists a time at which $c(V)$ relaxes to its equilibrium value. We call this equilibration time $V_{eq}$ that the system is at its equilibrium state. We define a time-dependent parameter
\be%
\label{equ}
\epsilon(V)=\bigg{|}\frac{c(V)-c_{eq}}{c(V)}\bigg{|},
\ee%
where the equilibration time is defined as the time which satisfies $\epsilon(V_{eq})<0.003$ and $\epsilon(V)$ stays below this limit afterwards. Note that $c_{eq}$ can be obtained from the static solution at the corresponding magnetic field and final temperature. This function has been plotted in figure \ref{thtime} for different values of the magnetic field. We can conclude from figure \ref{thtime} (right) that the presence of the magnetic field delays the evolution of the system to its equilibrium state. The equilibration times corresponding to the figure \ref{thtime} are given in table \ref{tableteq} which confirm, quantitatively, the previous statement.

\begin{table}[ht]
\label{tableteq}
\caption{Equilibration Times with respect to the External Magnetic Field}
\vspace{.5cm}
\centering
\begin{tabular}{c c }
\hline\hline
$B$ \qquad \qquad & \qquad \qquad $V_{eq}$   \\[0.5ex]
\hline
0.0 \qquad \qquad & \qquad \qquad  3.27834\\
0.3137 \qquad \qquad & \qquad \qquad  3.33708\\
0.5645 \qquad \qquad & \qquad \qquad  3.49866 \\
0.7129 \qquad \qquad & \qquad \qquad  4.8057\\
0.7946 \qquad \qquad & \qquad \qquad  5.15314\\[1ex]

\hline
\end{tabular}
\end{table}

%\section{Concluding Remarks}
%\newpage
\noindent\textbf{Acknowledgment}

\noindent We would like to thank T. Ishii, S. Kinoshita, K. Murata and N. Tanahashi for fruitful comments.

\end{document}